\documentclass{llncs}

\usepackage{hyperref}
\usepackage[hyphenbreaks]{breakurl}
\usepackage{paralist}
\usepackage{graphicx}
\usepackage[final]{pdfpages}
\usepackage{changebar}
\usepackage{epstopdf}
\usepackage[caption=false,font=footnotesize]{subfig}
\usepackage{rotating}	
\usepackage{multirow} 
\usepackage{arydshln}

\author{Daniel M\'{e}ndez Fern\'{a}ndez$^1$ and Stefan Wagner$^2$}

\institute{
Technische Universit{\"a}t M{\"u}nchen, Germany\\
\url{http://www4.in.tum.de/~mendezfe}
\and
University of Stuttgart, Germany\\
\url{http://www.iste.uni-stuttgart.de/}
}

\begin{document}

\title{A Case Study on \\Artefact-based RE Improvement in Practice}

\maketitle

\begin{abstract}

\emph{Background:} Most requirements engineering (RE) process improvement approaches are solution-driven and activity-based. They focus on the assessment of the RE of a company against an external norm of best practices. A consequence is that practitioners often have to rely on an improvement approach that skips a profound problem analysis and that results in an RE approach that might be alien to the organisational needs. 
\emph{Objective:} In recent years, we have developed an RE improvement approach (called \emph{ArtREPI}) that guides a holistic RE improvement against individual goals of a company putting primary attention to the quality of the artefacts. In this paper, we aim at exploring ArtREPI's benefits and limitations. 
\emph{Method:} We contribute an industrial evaluation of ArtREPI by relying on a case study research. 
\emph{Results:} Our results suggest that ArtREPI is well-suited for the establishment of an RE that reflects a specific organisational culture but to some extent at the cost of efficiency resulting from intensive discussions on a terminology that suits all involved stakeholders. 
\emph{Conclusions:} Our results reveal first benefits and limitations, but we can also conclude the need of longitudinal and independent investigations for which we herewith lay the foundation.

\end{abstract}

\section{Introduction}

Requirements engineering (RE) constitutes an important success factor for software development projects since stake\-holder-appropriate requirements are important determinants of quality. Its interdisciplinary nature, the uncertainty, and the complexity in the process, however, make the discipline difficult to investigate and to improve~\cite{MW14}. For an RE improvement, process engineers have to decide whether to opt for \emph{problem orientation} or for \emph{solution orientation}~\cite{PIGO08,NMJ09}. In a solution-driven improvement, the engineers assess and adapt their RE reference model, which provides a company-specific blueprint of RE practices and artefacts, against an external norm of best practices. The latter is meant to lead to a high quality RE based on universal, external goals (see, e.g.\ CMMI for RE~\cite{BHR05}). Solution-driven improvement approaches might thus serve the purpose of achieving externally predefined goals by implementing a set of best practices adhered by many organisations~\cite{PIGO08} (e.g.\ as part of a certification). They do not necessarily consider company-specific goals, however, that dictate the notion of RE quality within a particular socio-economic context (e.g.\ a company) and, thus, may result in an RE reference model that is alien to the organisational culture. In consequence, those RE improvement approaches encounter problems and are often rejected by practitioners~\cite{SNJA+07,MW14}. A notion of RE quality where company-specific goals dictate the improvement is the core of problem-driven approaches. 

Besides the improvement principles, the paradigm in which the targeted RE reference model is structured (and, thus, improved) plays an important role. A reference model can either be \emph{activity-based} or \emph{artefact-based}~\cite{MPKB10}. In short, an activity-based improvement approach focuses on improving the quality of the RE practices while an artefact-based one puts its focus on improving the quality of the RE artefacts. 

Most available RE improvement approaches today are solution-driven~\cite{MOWD14}. Yet, RE is complex by nature, and we postulate that RE quality depends on the contribution of an RE reference model to context-specific goals. Therefore, improvements cannot be meaningfully implemented without a qualitative problem investigation that reveals which goals must be achieved~\cite{NMJ09,PIGO08} and which artefacts should be created in which way. In response to the lack of problem-driven and artefact-based RE improvement approaches, we elaborated such an approach~\cite{MW2013} which we call \emph{ArtREPI}. We further realised our approach using the EPF Composer\footnote{\url{http://www.eclipse.org/epf/}} and made first experiences using ArtREPI in practice (see also Sect.~\ref{sec:RelatedWork}).

\textbf{Problem.} Although we made first conceptual and empirical contributions to support a problem-driven and artefact-based RE improvement, we still have little evidence on its practical benefits and limitations.

\textbf{Contribution.} In this paper, we report on the first industrial case study to evaluate \emph{ArtREPI} in comparison to solution-driven and activity-based REPI approaches previously used in the same contexts. The purpose, however, is not to evaluate only the particularities of our approach itself, but also 
\begin{compactenum}
\item to reveal first qualitative insights into the benefits and limitations of a problem-driven and artefact-based RE improvement in general, and 
\item to lay the foundation for future independent empirical investigations.
\end{compactenum}

Based on our contribution and the disclosed material~\cite{onlineArtREPI}, practitioners can therefore already apply our RE improvement approach, and researchers can build their conceptual and empirical work on our results to further explore the full spectrum of an RE improvement.

\textbf{Outline.} In Sect.~\ref{sec:RelatedWork}, we discuss fundamentals and related work. In Sect.~\ref{sec:ArtREPI}, we introduce the artefact-based RE improvement approach. We provide our case study design in Sect.~\ref{sec:CaseStudyDesign}, the results in Sect.~\ref{sec:CaseStudyResults}, and a critical reflection in Sect.~\ref{sec:Discussion}, before concluding our paper in Sect.~\ref{sec:Conclusion}.

\section{Fundamentals and Related Work}
\label{sec:RelatedWork}
Requirements engineering process improvement, as software process improvement in general, is a cyclic approach to continuously analyse problems/the current situation in RE as part of an appraisal, plann an improvement, realise the improvement and evaluate the improvement before initiating the next iteration. We can distinguish solution-driven approaches and problem-driven approaches as well as artefact and activity orientation.  

In literature, there exist mostly solution-driven contributions~\cite{NMJ09}. R-CMM, proposed by Beecham et al.~\cite{BHR05}, is a prominent representative of these approaches. It is based on CMMI and an empirical investigation in twelve companies~\cite{beecham2003spi}. The investigation revealed patterns and best practices based on problems experienced by practitioners. Therefore, it aimed at a generalised, external notion of RE quality. A technical validation using an expert panel~\cite{BHB+05} further illustrates selected success criteria, such as understandability. Approaches of this category focus on a solution-driven benchmarking of the maturity of RE according to a specific norm of best practices and may thus lead to the problems described in the introduction (see also~\cite{SNJA+07,MW14} for richer investigations).

In response to their shortcoming, Pettersson et al.\ contributed an approach to problem-driven RE improvement~\cite{PIGO08} called the iFLAP approach. Same as in ArtREPI, they make use of qualitative methods for the problem analysis and postulate the importance of strong stakeholder involvement. Although their concepts are promising to conduct a problem-driven REPI, the consequential next steps,  i.e.\ the actual improvement realisation by crafting a new RE reference model, was not in scope of their contribution.

To the best of our knowledge, there exists no holistic approach to a REPI covering all improvement phases in a seamless manner, let alone considering an improvement specifically directed at the RE artefacts. Recent work in this direction is made by Kuhrmann et al.~\cite{kb2014a}, but taking more a perspective on the management of software process models. The focus is thereby set on how to manage an artefact-based improvement rather than on how to conduct it. They do not look at how to analyse, design, and evaluate a process in a problem-driven manner focusing on the quality of the artefacts which is in scope of ArtREPI.
 
Available validation and evaluation research, which would be directly related to the contributions of this paper, focuses on the evaluation of methods or metrics used in isolated REPI phases, such as the analysis, on experience reports, or on the analysis of general success factors and lessons learnt. In the case study we present here, we therefore do not discuss the relation to existing evidence taking into account particular approaches, but exclusively take a qualitative view and rely on the evaluation of ArtREPI against the general perceptions and the experiences of the participants with solution-driven, activity-based REPI in the same context. Further details on the publication landscape on REPI can be taken from our previously published mapping study~\cite{MOWD14}.

\textbf{Previously Published Material.} In~\cite{MW2013}, we first introduced the basic concepts of ArtREPI and its design science principles. Since then, we realised our approach using the EPF Composer as a means of a technical validation and made all material (models, process documentation, document templates, and evaluation instruments) publicly available~\cite{onlineArtREPI} to support the dissemination. In a previous short paper~\cite{M2014}, we then briefly reported on initial experiences from an ongoing case study. In the paper at hands, we report on the by now completed case study in detail including the case study design, the results containing a second case, and the implications the results have.

\section{ArtREPI: Artefact-based RE Process Improvement}
\label{sec:ArtREPI}

\begin{figure}[htb]
\centering
  \includegraphics[width=0.95\textwidth]{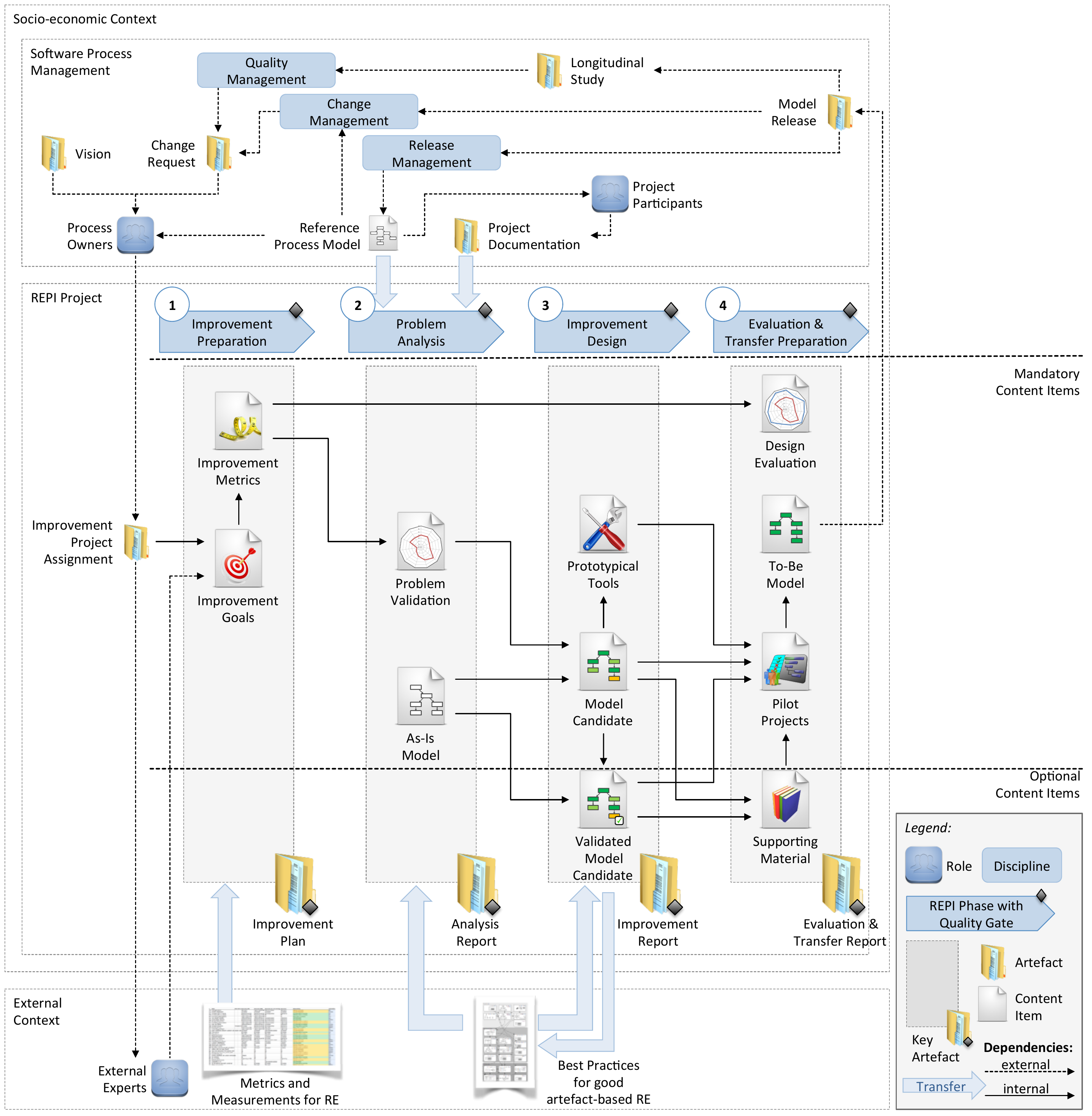}\\
 \caption{ArtREPI: Structuring and context.}\label{fig.ArtREPI_Context}
\end{figure}

Figure~\ref{fig.ArtREPI_Context} gives an overview of the basic structure of ArtREPI, which we use for our evaluation presented in this paper. 
As shown in the figure, we distinguish two contexts important to the notion of RE quality: an external context that contains norms of best practices and the socio-economic context where the notion of quality is dictated by individual demands. External norms of best practices are not only key to solution-driven RE improvement approaches but also important to a problem-driven improvement as one principle is to support technology transfer according to context-specific goals. 

The socio-economic context is further characterised by a set of disciplines that aim at managing software processes. Of particular interest is that we have process owners who usually have the sovereignty over an RE reference model (although often underrepresented in practice~\cite{MW14}) and a set of project participants who work according to an (explicitly established or implicitly lived) RE reference model. The latter is subject to an improvement in case an organisational change is triggered. For reasons of complexity, we omit the discussion of further disciplines or roles, such as an improvement sponsor. A change is performed as part of an RE process improvement project which is in scope of ArtREPI. ArtREPI consists of four phases and yields a set of mandatory and optional results. The mandatory ones eventually serve a seamless improvement based on a unified underlying data model of the improvement results. The full model can be taken from our online sources~\cite{onlineArtREPI} while in the following, we provide a brief overview of the single phases.

\subsection{Improvement Preparation}
In the preparation phase, we lay the groundwork for the improvement and aim at getting an understanding about the application domain, such as the chemical sector, typical constructs and rules followed therein, and the terminology used. We discuss contemporary problems and the primary improvement goals which, in collaboration with the process owners, we refine to concrete metrics and measurements to evaluate the success of the improvement after completion. Where possible and reasonable, we reuse metrics from previously conducted ArtREPI.
The outcome of the preparation phase is an improvement plan that defines the concrete procedures and time schedules as well as resource allocations (cases and subjects) for subsequent steps including interviews or workshops. 

\subsection{Problem Analysis}
The second stage aims at discovering problems in the use of the current RE reference model independent of whether it is explicitly defined as a company standard or not. In case a company standard exists, we analyse it first  to get an overview of the basic artefacts, roles, and activities. We then conduct a document analysis in selected exemplary development projects and abstract from the contents to build an as-is model of the RE artefacts, milestones, and roles. We complement this syntactic analysis with a semantic analysis where we analyse, for example, linguistic defects in document templates. We use the defined as-is model to conduct a gap analysis where we compare the current state of practice with an external standard wich serves as a knowledge base. More precisely, we rely on \emph{AMDiRE} (\textbf{a}rtefact \textbf{m}odel for \textbf{d}omain-\textbf{i}ndependent \textbf{RE}) that defines a best practice norm for artefact-based RE (see also~\cite{MP14}). We use potential gaps from our analysis as candidates for further validation which we do via semi-structured interviews. There, besides asking about experiences and expectations, we ask the participants why certain contents identified as incomplete in comparison to AMDiRE have or have not been specified in their projects. We take their reasoning as our primary input to establish later on a tailoring profile for the artefact-based RE reference model. The aim of the whole analysis is not to benchmark the process against an external standard. Instead and according to the problem-driven nature of ArtREPI, it aims at getting a better understanding on potential improvements which then can only be validated by project participants. To this end, we trigger  a self-reflection via interviews by indicating only  possibilities for changes that eventually only they can judge upon.

\subsection{Improvement Design}

In the third stage, we conduct the actual realisation of the improvement based on the candidates previously identified during the gap analysis. To this end, we conduct a series of action research workshops where we build a new model candidate by deciding which content items to consider in the candidate (e.g. ``use case model'') before we then subsequently define the actual content model that abstracts from concrete modelling concepts used to specify the content items in a project. The latter serves to build a prototypical (modelling) tool or document templates, and to establish content-related dependencies to the artefacts of further development phases, such as of testing, serving the purpose of a process integration into the overall software process model. We enrich the artefact model with process elements (roles, milestones) and a tailoring profile that emerges from the interviews and that defines under which project circumstances to document certain content items or not. Depending on the improvement project complexity or criticality, we may perform a validation of the model candidate before entering the last stage. In this validation step, we validate the model via feedback gathered by domain experts not involved in the improvement workshops. 

\subsection{Improvement Evaluation and Transfer Preparation}

The success of a problem-driven improvement can eventually only be determined by the degree to which the improvement outcome satisfies the improvement goals. In the last step, we therefore conduct technical action research workshops with project participants in  pilot projects where we apply the resulting RE reference model under realistic conditions. We specify a set of RE artefacts following the new RE reference model. Afterwards, we compare the outcomes and the followed process with the artefacts previously created in the same project environment by an interview. The participants rate the new model following the criteria agreed on in the preparation phase. In case of a positive rating, we release the new model complemented with supporting material.

\section{Case Study Design}
\label{sec:CaseStudyDesign}

In~\cite{MW2013}, we introduced in detail how we inferred our Art\-REPI approach presented in Sect.~\ref{sec:ArtREPI} from fundamental and applied research projects where we conducted an artefact-based and problem-oriented RE improvement. We argued so far that our approach is successful, because it emerged from successful RE improvements leading to new RE company standards. We were aware, however, that we need to better understand the benefits and limitations we can expect from ArtREPI and whether our approach can be used by others if we are not involved. This motivates the design for our empirical investigations:
\begin{compactenum}
\item We rely on \emph{case study research} with action research components in an industrial environment as we are particularly interested in elaborating qualitative insights into benefits and shortcomings in applying an RE improvement approach that also depends on subjective (and social) factors. The design follows the guideline of Runeson and H\"ost~\cite{runeson09}.
\item We included \emph{two} cases where the second one was conducted with little direct involvement of the ArtREPI authors. This should allow us to get an indication whether the success of an improvement depends on our influence, and it should provide a first step in scaling up to practice. 
\end{compactenum}

\subsection{Objectives and Research Questions}
\label{sec:RQ}

The study has the general objective to understand the benefits and limitations of applying ArtREPI in practice. 
We formulate two research questions to steer the study design structured into the evaluation of the improvement tasks (ArtREPI stage 1-3) and of the resulting RE reference model (ArtREPI stage 4):
\begin{compactitem}
\item[\textbf{RQ 1}] How well are process engineers supported in their RE improvement tasks? 
\item[\textbf{RQ 2}] How well are project participants supported by the resulting RE reference model?
\end{compactitem}

\subsection{Cases and Subjects}
\label{sec:StudyContext}
We applied ArtREPI as part of two improvement projects in two different contexts. For each project, we give a summary of the most important context information we are able to provide within the limits of existing non-disclosure agreements in Tab.~\ref{tab:ArtREPICases}. We chose the contexts because of their suitability to our research questions. We looked at two different settings (large versus small and agile). Furthermore,  REPI project 2 has, in contrast to the first case, only little involvement of those researchers developing ArtREPI (see also the next section). 

\begin{table}[bth]
\scriptsize
\centering 
\caption{Overview of cases (REPI 1 and REPI 2).}
\label{tab:ArtREPICases}
\begin{tabular}{lllp{.8\linewidth}l}
\hline
\textbf{No.} & \textbf{Aspects} & \textbf{Descriptions}\\ \hline \hline
1 & Improvement goal & Integration of RE into quality management. \\
 & REPI context & 2 process engineers, 4 domain experts, 4 external improvement consultants. \\ 
 & & Effort: 8 PM including 13 workshops. \\ 
  & Pilot projects & 3 pilot projects, 8 participants. \\ \hline
2 & Improvement goal & Re-design of RE to support agility.\\
 & REPI context & 2 main process engineers (previously coached by improvement consultants).  \\ 
 & & Effort: approx. 6 PM.\\
 & Pilot projects & 3 project participants to rate the reference model w/o pilot projects. \\ 
\hline
\end{tabular}
\end{table}

\textbf{REPI 1: Large scale RE process.} The first improvement project was conducted as part of a research cooperation between Technische Universit\"at M\"unchen and Wacker Chemie and was described, in parts, as the intermediate evaluation in~\cite{MW2013}.\footnote{In~\cite{MW2013}, we reported on first results from the first evaluation steps where the context-specific evaluation (ArtREPI stage 4) was still ongoing.} Wacker Chemie is a German company that works in the chemical business and develops custom software for their operation processes and their production sites. The improvement project aimed at defining a detailed artefact-based RE with a seamless integration into quality management. 

\textbf{REPI 2: Agile RE process.} The second improvement project was conducted as part of a Master's Thesis at the company SupplyOn AG which is a software as a service provider. The improvement project aimed at the re-design of the RE of the Rational Unified Process to an artefact-based agile RE approach. In contrast to project 1, where we evaluated the resulting RE reference model explicitly via pilot projects, we relied on informal reviews and expert opinion.

\subsection{Data Collection and Analysis Procedures}
We answer our research questions by collecting and analysing data from the application of ArtREPI and the assessments via questionnaires in both cases by improvement consultants who also authored ArtREPI. To this end, the improvement consultants coach the process engineers on ArtREPI, its underlying principles, and corresponding tools in a 3 hour workshop. These process engineers then apply ArtREPI as described in Sect.~\ref{sec:ArtREPI} and conduct the last phase of ArtREPI (i.e. the evaluation) by applying the resulting RE reference model in pilot projects of the company. In REPI 1, the improvement consultants formed part of the improvement project team after the coaching. In REPI 2, they where not directly involved anymore after the coaching sessions.

We collect the data for both research questions using questionnaires with open and closed questions. Each closed question is formulated as a statement where the participants should state their agreement on a Likert scale defined on an ordinal scale from 0 (\emph{``I strongly disagree''}) to a maximum of 7 (\emph{``I strongly agree''}) to avoid that they choose the middle. They have, however, always the possibility to refuse an answer. The open questions give them the possibility to provide a rationale for their decision. The original questionnaires can also be found in our online material~\cite{onlineArtREPI}. 

To answer RQ 1, we conduct an assessment by letting the involved process engineers rate ArtREPI on the basis of a questionnaire where they directly compare ArtREPI to previously followed solution-driven, activity-based improvement approaches. We summarise the questions in Tab.~\ref{tab:instrument2}. At the end, we finally asked the process engineers three concluding questions about the overall suitability of the approach to cover the particularities of their context.

\begin{table}[hbt]
\scriptsize
\centering 
\caption{Condensed RQ~1 instrument: Support of \textbf{process engineers} in improvement.}
\label{tab:instrument2}
\begin{tabular}{lp{9cm}}
\hline
\textbf{Criteria} & \textbf{Statements (closed question)} \\ \hline 
Structuredness & The approach was systematic.\\
Simplicity & The approach was easy to use.\\ 
Goal orientation & The approach considered problems and needs of all involved stakeholders. \\
Experience orientation & The approach considered company culture and stakeholder experiences.\\ 
Transparency & The decisions during the workshops were reproducible.\\ 
Effectivity & The improvement has led to the desired results.\\ 
Efficiency & I perceived the efficiency of the undertaking as high.\\ 
Knowledge transfer & The approach actively supported knowledge transfer.\\ \hline
Overall suitability (1) & ArtREPI was better suited than solution-driven approaches.\\
Overall suitability (2) & I would apply ArtREPI again. \\
Overall suitability (3) & I want to add following positive/negative aspects (open)\\
\hline
\end{tabular}
\end{table}

To answer RQ 2, we conduct an assessment by letting the involved project participants rate the outcome of ArtREPI using again a questionnaire. This evaluates the resulting RE reference model after application in the pilot projects in direct comparison with the one previously used in same context to evaluate whether the improvement goals have eventually been achieved.  Table~\ref{tab:instrument3} summarises the questions for this RQ. In the questions, we use ``model'' as a reference to the RE reference model (upper part in the table) and ``artefacts'' as a short reference for the artefacts created using the model (lower part in the table). 

\begin{table}[htb]
\scriptsize
\centering 
\caption{Condensed RQ~2 instrument: Support of \textbf{project participants} by RE reference model (resulting from the improvement tasks).}
\label{tab:instrument3}
\begin{tabular}{lp{8.5cm}}
\hline
\textbf{Criteria} & \textbf{Statements (closed question)} \\ \hline 
Flexibility & The model allows for flexibility in the RE process.\\
Ease of use & The model is easy to understand.\\ 
Effectivity & The application of the model has led to the desired results.\\ 
Efficiency & When applying the model, I perceived the efficiency as high. \\
Customisation / Tailoring & The model can be tailored to project-specific situation of the company.\\
Process integration & The model is integrated into further development activities and within the line organisation.\\ \hline 
Structuredness of artefacts & The artefacts are well structured and can be understood by people not involved in their creation.\\ 
Syntactic artefact quality & The model supports a high syntactic artefact quality w.r.t.  completeness and consistency.\\
Traceability of artefact & The model supports traceability within RE and between RE and further disciplines. \\
Semantic artefact quality & The model supports semantically consistent and complete artefacts.\\
Testability of artefacts & The model supports the creation of testable artefacts. \\ \hline
\end{tabular}
\end{table}

For the analysis of our results, we rely on descriptive analysis and qualitative interpretation of our data. We intentionally refrain from summarising visual accumulations of the ratings (e.g.\ via radar charts) because of the heterogeneity in the data as providing answers to each criterion was not obligatory. We therefore report on every rating given by each subject. We additionally indicate to whether the participants experienced an improvement or deterioration of applying ArtREPI in direct comparison to previously used improvement approach which we assume to be present if the mean values differ by at least one point.

\section{Case Study Results}
\label{sec:CaseStudyResults}

In the following, we summarise the results from the rating given in the assessments, structured according to the research questions. For reasons of confidentiality, we cannot provide details about workshop contents, the company-specific RE reference model and the pilot projects, but we will report the rating results.

\subsection{RQ~1: Support in RE Improvement Tasks}
Table~\ref{tab:resultsRQ2} summarises the ratings of the ArtREPI approach by the process engineers in direct comparison to previously used approaches which were based on CMMI. For each criterion, we show each subject's ratings. We had 6 subjects in context 1 and 2 subjects in context 2 (see~Tab.\ref{tab:ArtREPICases}). Furthermore, we give the mean and median as central tendencies and further show whether ArtREPI is clearly considered better ($+$) or worse ($-$). No clear comparison is indicated by a $0$. At the bottom of the table, we show the results of the rating of whether ArtREPI was considered better suited for the respective context than previously used REPI approaches.

\begin{table*}[!htb]
\scriptsize
\centering
\caption{Results for RQ~1: Rating of improvement procedure by subjects $S_{x}$ on a scale of 0 (\emph{``I strongly disagree''}) to 7 (\emph{``I strongly agree''}) from a \textbf{process engineering perspective}. See also Tab.~\ref{tab:instrument2} for details on the used instrument.}
\label{tab:resultsRQ2}
\begin{tabular}{lc  cccccc  ccc c cc  ccc}
                                                  &                    & \multicolumn{9}{c}{REPI 1}    &                                                    & \multicolumn{5}{c}{REPI 2 } \\ \cline{3-11} \cline{13-17}
                       &  		     & S1  & S2  & S3 & S4 & S5 & S6 & Mean & Median & Comp.  &                  & S1 & S2 & Mean & Median & Comp. \\ \hline 
\multirow{2}{*}{Structuredness} & \emph{Old} & - & - & 5  & 6  & 5  & 3  & 4.75 & 5   & \multirow{2}{*}{$+$} & &  6  &  4  &   5  &  5   &   \multirow{2}{*}{0}   \\ 
                               		        & \emph{New}& 6    & 5     & 6  & 7  & 7  & 5  & 6 & 6    &                              &          &  6  &  5  &  5.5   &   5.5&    \\ 

\multirow{2}{*}{Simplicity} & \emph{Old} & - & - & 5  & 5  & 5  & 5  & 5 & 5   & \multirow{2}{*}{0} & &  6  &  5  &   5.5  &  5.5   &   \multirow{2}{*}{0}   \\ 
                               		        & \emph{New} & 5    & 5     & 6  & 5  & 5  & 3  & 4.83 & 5    &  &                                      &  6  &  5  &  5.5   &   5.5 &    \\ 
		        
\multirow{2}{*}{Goal Orientation} & \emph{Old} & - & - & 4  & 6  & 4  & 4  & 4.5 & 4   & \multirow{2}{*}{0} & &  4  &  4  &   4  &  4   &   \multirow{2}{*}{$+$}   \\ 
                               		        & \emph{New} & 3    & 3     & 5  & 6  & 6  & 4  & 4.5 & 4.5    & &                                       &  6  &  6  &  6   &   6 &    \\ 
		        
\multirow{2}{*}{Experience Orientation} & \emph{Old} & - & - & 4  & -  & 3  & 6  & 4.33 & 4   & \multirow{2}{*}{$-$} & &  6  &  6  &   6  &  6  &   \multirow{2}{*}{$-$}   \\ 
	                                 		        & \emph{New} & 2    & 2     & 5  & -  & 5  & 2  & 3.2 & 2    & &                                       &  4  &  3  &  3.5   &   3.5 &    \\ 

\multirow{2}{*}{Sustainability} & \emph{Old} & - & - & 5  & 4  & 4  & 4  & 4.25 & 4   & \multirow{2}{*}{$+$} & &  4  &  5  &   4.5  &  4.5   &   \multirow{2}{*}{0}   \\
	                                 		        & \emph{New} & 6    & 5     & 6  & 4  & 7  & 4  & 5.33 & 5    &               &                         &  4  &  5  &  4.5   &   4.5 &    \\ 

\multirow{2}{*}{Effectivity} & \emph{Old} & - & - & 5  & 5  & 6  & 3  & 4.75 & 5   & \multirow{2}{*}{0} & &  3  &  4 &   3.5  &  3.5   &   \multirow{2}{*}{$+$}   \\ 
	                                 		        & \emph{New} & 3    & 3     & 6  & 5  & 6  & 5  & 4.66 & 5    &  &                                      &  7  &  6  &  6.5   &   6.5 &    \\ 

\multirow{2}{*}{Efficiency} & \emph{Old} & - & - & 5  & 6  & 5  & 4  & 5 & 5   & \multirow{2}{*}{$-$} & &  3  &  4 &   3.5  &  3.5   &   \multirow{2}{*}{$+$}   \\ 
	                                 		        & \emph{New} & 1    & 2     & 6  & 6  & 5  & 4  & 4 & 4.5    &   &                                     &  7  &  5  &  6   &   6 &    \\ 

\multirow{2}{*}{Knowledge Transfer} & \emph{Old} & - & - & 4  & 6  & 4  & 4  & 4.5 & 4   & \multirow{2}{*}{0} & &  3  &  4 &   3.5  &  3.5   &   \multirow{2}{*}{$+$}   \\ 
	                                 		        & \emph{New} & 5    & 5     & 5  & 6  & 6  & 3  & 5 & 5    &           &                             &  7  &  6  &  6.5   &   6.5 &    \\ \hline 
Overall Suitability  & \emph{New} & - & - & 5  & -  & -  & 5  & 5 & 5   &  &  & 7  &  6 &   6.5  &  6.5   &   \\ \hline
			        
\end{tabular}
\end{table*}

Overall, ArtREPI was rated as a structured improvement approach that tends to better support knowledge transfer than previously used approaches. Surprising to us, however, is the result of REPI 2 conducted by people not involved at all in the development of ArtREPI. Our assumption was that the rating would be worse than in REPI 1, but it was better regarding the goal orientation, the effectivity and efficiency, and the support for knowledge transfer. Qualitative statements from the open questions provide some explanations. The subjects in REPI 1 rated that the initial preparation phase was performed in a ``too academic'' fashion with ``too many discussions to clarify the terminology'', especially by those subjects with no REPI experience made before (S1 and S2). This might also be the reason for the negative comparison regarding the experience orientation. In contrast, REPI 2 was conducted solely by process engineers employed by the respective company and familiar with the culture and the domain.

Finally, the overall rating whether ArtREPI was better suited to achieve the company-specific improvement goals in comparison to previously used approaches were answered positively in both cases. The process engineers stated for both cases that they would apply ArtREPI in follow-up improvement cycles. The answers to the open questions showed for REPI 1, however, that the engineers expect an integration into the organisation as a prerequisite for a repetition. Another suggestion was to check in a follow-up study whether the efficiency of ArtREPI could be improved. Further informal statements included that: 
\begin{compactenum}
\item the action research workshops, where the new artefact-based RE reference model was crafted by the process engineers, together with the group discussions, fostered discussions the engineers would otherwise not have, e.g. about roles and responsibilities in the RE that just seemed clear to everybody.
\item independent of the results from the pilot projects (described next), they would need additional longitudinal studies before declaring the new RE reference model as a new standard as too many changes have been made. 
\end{compactenum}

\subsection{RQ~2: Support by resulting RE Reference Model}

The success of an improvement eventually depends on whether the resulting RE reference model achieves the improvement goals. RQ~2 therefore focuses on evaluating the REPI outcome in pilot projects. Table~\ref{tab:resultsRQ3} shows the rating of the artefact-based RE reference model resulting from the improvement cycle from the perspective of project participants. We distinguish criteria to rate the application of the reference model itself and criteria to evaluate the RE artefacts produced following the reference model. In REPI 1, we applied the model in three pilot projects (two with custom software development and one with standard software). In REPI 2, it was not possible to fully implement the new model immediately, but they included the artefacts created during the coaching sessions. Therefore, their rating is does not reflect the full experience with ArtREPI. 
\begin{table*}[hbt]
\scriptsize
\centering
\caption{Results for RQ~2: Rating of resulting RE reference model by subjects $S_{x}$ on a scale of 0 (\emph{``I strongly disagree''}) to 7 (\emph{``I strongly agree''}) from a \textbf{project perspective}.}
\label{tab:resultsRQ3}
\begin{tabular}{llc  cccccccc  ccc c ccc  ccc} 
&                                                                &               & \multicolumn{11}{c}{REPI 1}		&		                                                              & \multicolumn{6}{c}{REPI 2} \\ \cline{4-14} \cline{16-21}
&                                     & & S1  & S2   & S3    & S4 & S5 & S6   & S7 & S8  & Mean & Med. & Comp.       &    				   & S1 & S2 & S3  & Mean & Med. & Comp. \\ \hline 

\multirow{12}{*}{\begin{sideways}Process Quality\end{sideways}} & \multirow{2}{*}{Flexibility} & \emph{Old}   & 7    & 7      & 0  & 4    & 2    & 5     &  6  & 6    & 4.63         & 5.5        & \multirow{2}{*}{0}   &    &   5  &  6  &   1  & 4        & 5       &    \multirow{2}{*}{$+$}   \\ 
&                                          & \emph{New}& 4    & 4      & 6      & 6   & 5     & 2     & 6   & 5    & 4.75        & 5         &    &                                  		   &  6   &  6  &  7   &6.33         & 6       &    \\ 
                                          
&\multirow{2}{*}{Ease of Use} & \emph{Old}   & 5    & -      & 5  & 4    & 6    & 4     &  4  & 5    & 4.71         & 5        & \multirow{2}{*}{$-$}  & &   5  &  2  &   5  &4          & 5        &    \multirow{2}{*}{$+$}   \\ 
  &                                        & \emph{New}& 3    & -      & 5      & 3   & 5     & 2     & 4   & 4    & 3.71        & 4         &                                    &  		   &  5   &  5 &  5   &5          & 5       &    \\ 

&\multirow{2}{*}{Effectivity} & \emph{Old}   & 4    & -      & 5  & 6    & -    & 5     &  6  & 4    & 5         & 5        & \multirow{2}{*}{0}  & &   5  &  -  &   1  &3          & 3        &    \multirow{2}{*}{$+$}   \\ 
  &                                        & \emph{New}& 6    & -      & 5      & 6   & -     & 2     & 6   & 6    & 5.16 & 6         &                               &       	   &  6   &  -  &  7   &6.5          & 6.5       &    \\  

&\multirow{2}{*}{Efficiency} & \emph{Old}   & 3    & -      & 4  & 5    & -    & 6     &  6  & 5   & 4.83         & 5        & \multirow{2}{*}{$-$} &   &   6  &  -  &   0  & 3          & 3       &    \multirow{2}{*}{$+$}   \\ 
  &                                        & \emph{New}& 1    & -      & 5      & 6   &    -  & 2     & 6   & 3    & 3.83        & 4         &                              &                  &  5   & -   &  7   &6          & 6      &    \\  

&\multirow{2}{*}{Customisation} & \emph{Old}   & 7    &    -   & 1  & 3    & 1    & 6     &  6  & 4    & 4         & 4        & \multirow{2}{*}{0}    &   &   5  &  7  &   4  &5.33          &     5    &    \multirow{2}{*}{$+$}   \\ 
  &                                        & \emph{New}        & 6    &   -    & 6   & 6   & 4     & 1     & 6   & 5    & 4.86        & 6        &                              &        		   &  6   &  7 &  7   &6.67          & 7       &    \\  

&Process & \emph{Old}   & 3    &   -    & 2  & 3    & 3    & 6     &  5  & 3    &3.75         & 3        & \multirow{2}{*}{$+$}   &    &   5  &  3  &   5  & 4.33          & 5        &    \multirow{2}{*}{$+$}   \\ 
&Integration                                          & \emph{New}& 6    &    -   & 4      & 4   & 6     &6     & 5   & 5    & 5.14        &    5  &    &                                      		   &  5   &  6  &  5   &5.33          & 5       &    \\ \hline 
                       
\multirow{10}{*}{\begin{sideways}Artefact Quality\end{sideways}} & \multirow{2}{*}{Structuredness} & \emph{Old}   & 2    & 5      & 3  & 5    & 3    & 2     &  5  & 2    & 3.38         & 3        & \multirow{2}{*}{0}  &     &   3  &  4  &   2  & 3          &     3    &    \multirow{2}{*}{$+$}   \\ 
  &                                        & \emph{New}          & 4    & 7      & 3      & 6   & 4     & 2     & 5   & 3    & 4.25        & 4         &                                      		 &  &  3   &  6  &  6   &5          & 6       &    \\  

& Syntactic & \emph{Old}   &  0  & 5      & 4  & 4    & 0    & 2     &  6  & 3    & 3         & 3.5        & \multirow{2}{*}{$+$}   &    &  -  &  2  &   2  &2          & 2        &    \multirow{2}{*}{$+$}   \\ 
& Quality                             & \emph{New}& 7    & 7      & 5      & 5   & 2     & 5     & 6   & 6    & 5.38        & 5.5         &    &                                  		   &  -   &  7  &  6   &6.5          &      6.5  &    \\  

&\multirow{2}{*}{Traceability} & \emph{Old}   & 4    & 4      & 1  & 3    & 2    & 2     &  6  & 3    & 3.13         & 3       & \multirow{2}{*}{$+$}   &    &   1 &  3  &   0  & 1.33          &    2     &    \multirow{2}{*}{$+$}   \\
 &                                         & \emph{New}   & 7    & 6      & 6      & 7   & 4     & 5     & 6   & 4    & 5.63        & 6         &                                 &     		   &  6  &  6  &  6   &6          & 6       &    \\  

&Semantic & \emph{Old}   & 4    & 4      & 2  & 3    &  -   & 3     &  7  & 3    & 3.71         & 3        & \multirow{2}{*}{$+$}  &     &   4 &  3  &   5  &4          &     4   &    \multirow{2}{*}{$+$}   \\ 
&          Quality                           & \emph{New}& 6    & 6      & 5      & 4   & -     & 5     & 7   & 6    & 5.57        & 6        &   &                                   		   &  6   &  6  &  5   &5.67          & 6       &    \\  
                                          
&\multirow{2}{*}{Testability} & \emph{Old}   & 1    & 2     & 3  & 4    & 5    & 2     &  6  & 3    & 3.25         & 3        & \multirow{2}{*}{$+$} &     &   5 &  3  &   3  &3.67          & 3        &    \multirow{2}{*}{$+$}   \\ 
 &                                         & \emph{New}& 5    & 4      & 5      & 6   & 7     & 5     & 6   & 5    & 5.38        & 5         &                                &      		   &  6   &  7  &  6   &6.33          & 6       &    \\  \hline
\end{tabular}
\end{table*}

The results indicate that ArtREPI supported the participants in achieving their improvement goals (see Sect. \ref{sec:StudyContext}). For REPI 1, we could improve, e.g., the traceability, the testability, and the process integration of the RE reference model, thus, supporting the better integration into quality management which formed the improvement goal. The negative results in the comparison to the previously used RE reference model regarding the process quality might be explained from the complexity of the new (richer) model and the learning curve associated with all new methods in general. In contrast, the positive rating in REPI 2 might be explained by the new RE reference model to support agility due to its new (light weight) simplicity. 

Overall, we observe that ArtREPI can show its strength especially in the artefact quality as expected, because it is oriented towards artefacts. In the process quality, the picture is, however, more mixed. Qualitative feedback to the open questions provides explanations:
\begin{compactitem}
\item The success of a pilot study strongly depends, beyond political factors, on the quality of the prototypical implementation. Hence, project participants should be involved in the technical validation, too.
\item Coaching sessions before the pilot studies should not only focus on the new RE reference models, but on the underlying principles, e.g.\ new levels of abstraction in the requirements or new roles and responsibilities.
\item Although artefact-based RE reference models are inherently process agnostic, they should include suggestions for methods and modelling techniques (e.g. UML-based ones) project participants are familiar with. This should increase the organisational willingness to change.
\end{compactitem}

\subsection{Threats to Validity}
\label{sec:Threats}
There exist many threats to validity inherent to case study research~\cite{runeson09}. Threats to the \emph{internal validity} mainly arise from potential bias during the data collection, let alone because of our action research components, and because of the general subjective nature in the ratings. We applied selected techniques to reduce those threats, e.g. researcher and method triangulation, but to a certain extent we were particularly interested in gathering subjective opinions by the interviewees. Another mitigation strategy followed in advance was to apply ArtREPI in the second case with little influence of the approach authors. It provides a first indication towards scaling up to practice~\cite{Wieringa13}, thus, it provides a first step in strengthening the \emph{external validity}. This generalisation, however, needs further attention in future work. 

Finally, a more general problem is that many things remain (objectively) unmeasurable. As a matter of fact, we still have a limited understanding on how to reliably measure long-term improvement effects going beyond RE, because of the complexity of confounding variables in a software project ecosystem. In our studies, we therefore refrained from such measurements in advance. Measurements are still important to determine the success of an improvement given that (1) RE forms part of a larger context that needs to be taken into account, while (2) we consider problem orientation where we cannot rely on an external reference to determine a notion of software process quality (in the sense of an oracle). 

Another facet important to an improvement is finally the question how much the notion of process quality is determinable by the quality of the artefacts and also how much project participants eventually rely on the created artefacts. We can observe first empirical investigations in that direction (e.g.~\cite{L15}). Still, even if we can measure certain phenomena on basis of the artefacts, we still do not fully understand to which extend the notion of RE quality eventually manifests itself in the created artefacts. That is, the investigations are based on the critical assumption that the success of a project and, in particular, of an artefact-based improvement depends on the documented results on which project participants rely. To elaborate the extent to which the application of ArtREPI eventually leads to an improvement, and how to measure the success of an improvement (including subjective and cognitive facets), we first need a better understanding on the measurability of such an improvement. In~\cite{MMFV14}, we provide a richer discussion on the limitations of measurements in RE. 

\section{Discussion}
\label{sec:Discussion}

Our results indicate that  ArtREPI is well suited to cover the needs of a structured improvement where problem and artefact orientation are important, while supporting knowledge transfer by continuous stakeholder involvement. The direct comparison of the two cases further indicates that the effectivity, efficiency, the knowledge transfer, and even the goal orientation are strengthened when the improvement is conducted by company members with no intervention from outside. The results from pilot projects further suggest that the improvement eventually achieved the local improvement goals. Our overall results therefore strengthen our confidence in the benefits of ArtREPI as a self-contained and holistic approach to a problem-driven RE improvement as long as the improvement goals are in tune with the (known) benefits of artefact orientation. 

\subsection{Limitations of ArtREPI}
One benefit in case study research is that it gives us the possibility to get qualitative feedback, i.e.\ explanations for the particular ratings. We are particularly interested in revealing limitations of ArtREPI as this supports us steering the development and the evaluation of the discussed improvement principles, and it helps us fostering the discussions on RE improvement in general. 

Limitations we could reveal by our second case concern more social aspects of a process improvement. For example, we, as researchers involved in the development of ArtREPI, seemed to lower the efficiency and effectivity of an improvement due to long preparation phases to increase our understanding of the domain and the terminology used. Our initial assumption that the success of ArtREPI depends on that we ourselves should conduct the improvement thus is wrong. However, the exact consequences of applying ArtREPI without any involvement of the authors at all (e.g.\ without coaching) remain still unknown.

Also, there exists a plethora of organisational factors, such as the support by the management, the support at project level (a champion), and general social as well as empirical skills that constitute success factor for an improvement. It might be possible to include some factors in the approach. For instance, one suggestion was to involve project participants earlier during technical validation stages to mitigate the threats arising from a missing organisational willingness to change. Other factors, however, might not be covered at all, because models abstracts, by nature, from desires, beliefs, experiences, and expectations, which all are important. To fully reveal those factors, and to fully explore to which extent they can eventually be influenced on a methodological layer, we need more (especially longitudinal) investigations in that direction.

\subsection{Success Factors for ArtREPI}
Besides the general focus of problem orientation, which is to emphasise context-specific goals over the possibility for an external certification of an RE, we found several factors that influence the success of an RE improvement:

\begin{compactitem}
\item Improvement goals need to be in tune with the expected benefits of the chosen paradigms (in our case artefact orientation) known in advance.\footnote{This implies that we need sufficient evidence on potential benefits of artefact orientation to the chosen context beforehand to justify the decision to conduct an organisational change.}
\item Support by higher management, especially when communicating new roles and responsibilities.
\item Backup by project environments, e.g. via the early involvement of project participants in the validation stages of a new RE reference model (before initiating the evaluation stages).
\item Domain knowledge. 
\item Reflection of the organisational culture.
\item Social and empirical skills.
\end{compactitem} 

Those factors might not be surprising in themselves but together in their extent. The success of a problem-driven improvement is determined by the reflection of the organisational culture of a company in every facet of the improvement approach. That is, all relevant stakeholders need to be involved in early stages to cover their needs, a backup by representative projects needs to be ensured (in the sense of champions), and project participants need to identify themselves with the resulting reference model. All those facets aim at supporting the willingness for a change in the way of working to reshape an existing organisation~\cite{Pfleeger99}.

In our current understanding, this can only be achieved by applying exhaustive, qualitative empirical method that foster continuous stakeholder involvement but which also come themselves with limitations, e.g.\ regarding the measurability of improvement effects (see also our Sect.~\ref{sec:Threats}).

\section{Conclusion}
\label{sec:Conclusion}

In this paper, we reported on our first steps to empirically evaluate the principles of an artefact-based and problem-driven RE improvement (ArtREPI). Our investigation should give a first qualitative impression of the general benefits and shortcomings in direct comparison to existing principles of solution-driven RE improvements that currently dominate the publication landscape~\cite{MOWD14}. The purpose was also to lay the foundation for independent empirical investigations to eventually explore the full spectrum of RE improvement.

To this end, we reported on two industrial case studies and analysed (1) how ArtREPI would support process engineers in achieving a problem-driven and artefact-based RE improvement, and (2) how well project participants were supported in their project environments by the resulting artefact-based RE reference model. We further conducted the second case with very little influence of the approach authors to investigate, in a first step, to what extent the results might be influenced by us researchers being involved in the RE improvement. 

Our results strengthen our confidence that ArtREPI is suitable for a self-contained, problem-driven RE improvement where the improvement goals are in tune with the scope of artefact orientation. More important, we could also reveal first limitations and factors important to future work in RE improvement research. We need to further explore the measurability of an improvement and its long-term effects, and to further support scaling up to practice. To support the generalisation of our observations that, so far, are only valid for chosen local improvement contexts~\cite{Wieringa13}, we need to (1) foster independent replications of our case studies which we support by our contribution and our publicly accessible online material~\cite{onlineArtREPI}, and (2) we need to take new perspectives. 

Rather than further investigating the technicalities of applying ArtREPI ourselves, we need to concentrate on exploring further social and cognitive facets and the long-term effects of an RE improvement for which our contribution has provided one foundation and sensitisation. We thereby encourage researchers and practitioners to join us in this endeavour to fully understand the broad spectrum of possibilities and limitations in an artefact-based and problem-driven RE improvement. 

\textbf{Acknowledgements.} We want to thank R.~Wieringa and M.~Broy for the collaboration during the development of ArtREPI. We further want to thank R.~Bossek and M.~Kuhrmann for their support during the systematisation of ArtREPI, and S.~Wiesi, J.~Mund, J.~Eckhardt, and H.~Femmer for their support in the REPI projects. Finally, we are grateful to all subjects involved during the case studies and to  B.~Penzenstadler, M.~Daneva, and A.~Vetr\`{o} for their valuable feedback on previous versions of this manuscript.

\bibliographystyle{splncs}
\bibliography{profes15}
\end{document}